\documentclass[preprint]{aastex}
\newcommand \kms{km~$\rm{s}^{-1}$}
\newcommand \cc{$\rm{cm}^{-3}$}

\newcommand \Ha{H$\alpha$}

\newcommand \mum{$\mu$m}
\newcommand \lsol{L$_{\odot}$}

\newcommand \fdens{erg s$^{-1}$ cm$^{-2}$ arcsec$^{-2}$}
\newcommand \flux{erg s$^{-1}$ cm$^{-2}$}
\newfont{\rten}{cmr10} 
\def \arcdeg{\hbox{$^\circ$}}
\def \arcmin{\hbox{$^\prime$}}
\def \arcsec{\hbox{$^{\prime\prime}$}}

\begin{document}
\normalsize

\title{HST NICMOS Images of the HH 7/11 Outflow in NGC1333$^1$}

\author{A. Noriega-Crespo\altaffilmark{1}, A. Cotera 
\altaffilmark{2}, E. Young\altaffilmark{2,},
and H. Chen\altaffilmark{2}}

\altaffiltext{1}{Based on observations made with the NASA/ESA 
{\it Hubble Space Telescope}, which is operated by the
Association of Universities for Research, Inc. and under NASA contract
NAS5-26555.}

\altaffiltext{1}{SIRTF Science Center, Caltech 100-22,
Pasadena, CA, 91125, USA}

\altaffiltext{2}{Steward Observatory, University of Arizona,
933 N. Cherry Av, AZ, 85721, USA}

\begin{abstract}
We present near infrared images in H$_2$ at 2.12\mum~of the HH 7/11 outflow
and its driving source SVS 13 taken with {\it Hubble Space Telescope}  
NICMOS 2 camera, as well as archival \Ha~and [\ion{S}{2}] optical
images obtained with the WFPC2 camera.
The NICMOS high angular resolution observations confirm the
nature of a small scale jet arising from SVS 13, and resolve a
structure in the HH 7 working surface that could correspond to
Mach disk H$_2$ emission. The H$_2$ jet has a length of 
430 AU (at a distance of 350 pc), an aspect ratio of 2.2 and
morphologically resembles the well known DG Tau optical micro-jet. 
The kinematical age of the jet ($\sim 10$ yr) coincides with 
the time since the last outburst from SVS 13.
If we interpret the observed H$_2$ flux density with 
molecular shock models of 20-30 \kms, then the jet has
a density as high as $10^5$ \cc. The presence of this 
small jet warns that contamination by H$_2$ emission from an outflow 
in studies searching for H$_2$ in circumstellar disks is possible.
At the working surface, the smooth H$_2$ morphology of the HH 7 bowshock 
indicates that the magnetic field is strong, playing a major role in 
stabilizing this structure. The H$_2$ flux density 
of the Mach disk, when compared with that of the bowshock, suggests that 
its emission is produced by molecular shocks of less than 20 \kms.
The WFPC2 optical images display several of the global features already 
inferred from groundbased observations, like the  filamentary structure
in HH 8 and HH 10, which suggests a strong interaction of the outflow with
its cavity. The H$_2$ jet is not detected in [\ion{S}{2}] or \Ha~, 
however, there is a small clump at $\sim 5$ \arcsec~NE of SVS 13 that could
be depicting the presence either of a different outburst event or the north
edge of the outflow cavity.
\end{abstract}


\section{Introduction}
\label{intro}

Herbig-Haro (HH) objects trace optically the mass loss process from 
young stellar objects (YSOs) and their interaction with the surrounding 
medium (Reipurth  \& Bally~\citeyear{bo01}). Because of their morphology, 
energetics and size, HH objects are an integral part of the star 
formation process and its effect on molecular clouds. 
HH 7/11 is a chain of HH objects in 
Herbig's photographic plate catalogue (Herbig~\citeyear{her74}),
located in the very active star forming region NGC 1333 
(Aspin et al.~\citeyear{ASR94}; Bally, Devine \& Reipurth~\citeyear{betal96})
at a distance of 350pc (Herbig \& Jones~\citeyear{her83}). 
From the ground, at optical wavelengths, the HH 7/11 
system is defined by an arch-shaped morphology (blue lobe) that spans 
$\sim$ 2\arcmin. The red-shifted counter-lobe is detected in the near 
infrared (NIR), e.~g. at 2.12\mum~in the H$_2$ (1,0) S(1) line,
and displays a more chaotic structure (Stapelfeldt et al.~\citeyear{sta91}; Garden, 
Russell \& Burton~\citeyear{gar90}). 
Recent interferometric observations  (Bachiller et al.~\citeyear{bach00}) 
have convincingly demonstrated that SVS 13 
(Strom, Vrba \& Strom~\citeyear{str76})  
is the driving source of the HH 7/11 outflow. SVS 13 has a Class 0/I 
spectral energy distribution (Bachiller et al.~\citeyear{bach98}) 
and a luminosity of $\sim 85$~\lsol~
(Molinari, Liseau \& Lorenzetti~\citeyear{mol93}).

Detailed optical spectroscopic observations of HH 7/11 
show a complex velocity field and a low excitation nature, consistent 
with shock velocities of 30-60 \kms~(Solf \& B\"ohm~\citeyear{sb87};
B\"ohm \& Solf~\citeyear{khb90}). NIR spectroscopy displays a rich H$_2$ 
vibrational spectra and strong [Fe~II] 1.257 and 1.644\mum~lines
(Gredel~\citeyear{gre96}; Everett~\citeyear{eve97}), 
again consistent with collisional excitation by shocks.
Far infrared spectroscopy, which includes the emission of molecular 
species (like H$_2$, H$_2$O and CO) and atomic fine structure lines 
(like [O~I] 63\mum~and [Si~II] 34.8\mum), requires both J-type and 
C-type shocks of 15-30 \kms~to explain their ratios 
(Molinari et al.~\citeyear{mol00}).

In this paper we present new high angular resolution (FWHM=0.1\arcsec) 
{\it Hubble Space Telescope} (HST) NICMOS H$_2$ images at 2.121\mum~of
 HH 7/11 and its source SVS 13, as well as archive optical 
images in \Ha~and [\ion{S}{2}] taken with the WFPC2 camera
at a similar epoch.

\section{Observations and Data Reduction}
\label{obs}

The NICMOS observations were made with Camera~2 which has a nominal 
plate scale of $0.''0755\pm0.''005$ per pixel.  
The observations were carried out on January 9, 1998.  Three filters 
were selected for observation, F187N (P$\alpha$), F204M, and F212N (H$_2$). 
Two dithered images were taken with the three filters sequentially 
at each of five pointing positions.  Integration times 
per frame were 40 sec at  F187N and F204M, and 80 sec for
F212N, resulting in a total integration time of 80 sec in F187N and F204M, 
and 160 sec in F212N. Dark frames used for subtraction were taken at the 
end of the observations.  

The images were reduced using the IRAF data reduction package NICRED, 
written for HST/NICMOS data by McLeod and Rieke~\citeyear{mcleod97}). 
Darks were created using the observed dark frames with the routine 
NICSKYDARK, part of the NICRED package. The flats were those 
produced by M. Rieke to be used with NICMOS data.  After dark subtraction 
and flat fielding, the images were cleaned for any additional bad pixels
using the IRAF routine IMEDIT. 

The NICMOS pixels are non-square by $\sim$1\%, and prior to shifting 
and adding, the pixels were rectified using IDL procedures developed 
for the Image Display Paradigm \#3 (IDP3) \footnote{IDP3 is publicly available 
at http://nicmos.as.arizona.edu/software/idl-tools/toollist.cgi}
software package (Stobie et al.~\citeyear{stobie99}). 
IDP3 was then used to shift the data, aligning them using the world 
coordinate system values, and median combine all the images.  
The combined images were then flux calibrated using the 
values derived by M. Rieke for NICMOS data 
(1999 private communication).  

Observations on the adjacent narrow band continuum filters to the 
F212N and F187N were not taken due to time constraints, so a method had 
to be developed to provide an useful empirical estimate of the continuum 
in the narrow band filters using the F204M continuum image. 
The continuum subtraction was performed interactively using IDP3.
Since the continuum image is taken with a much wider filter, 
and at a central wavelength not immediately adjacent to the narrow band
filter, they needed to be scaled prior to subtraction.  
In IDP3, the continuum image was precisely aligned with the emission line 
image using the SVS 13 star, and subtracted from the narrow band image with 
a slowly increasing scale factor until the variance in the differenced image 
of SVS 13 was minimized.

The WPFC2 observations were taken from the HST archive and reduced using 
the standard IRAF STSDAS packages. The optical images unfortunately do not
cover the entire HH 7/11 outflow; the brightest HH 7 object is partially 
missing at edge of one of the detectors. The lack of reference stars between 
the frames, implies that the detailed comparison relies on the
accuracy of the coordinates systems adopted by NICMOS and WFPC2.
The source SVS 13 was used to align the images, first by centroiding, then by 
minimizing the residuals of differenced images.  A summary of the observations
is in Table 1.

\section{Results and Discussion}

Before comparing the high resolution images from HST, let's
briefly revise what is observed from the ground to emphasize the
differences and similarities between the emission of the atomic gas and
the warm molecular Hydrogen, in particular because the WFPC2 images
do not cover the entire object. In Figure \ref{f1}, 
we compare two images of HH 7/11 placed on the same scale, one taken in 
[\ion{S}{2}] 6717/31 \AA~(from Noriega-Crespo \& Garnavich~\citeyear{nori01}),
and other taken in H$_2$ at 2.121\mum.
The H$_2$ image was obtained with the Fred L. Whipple (FLWO)
1.2 m telescope and the ADS SBRC Camera in 1997 on October 10, with
an angular resolution of $\sim 1.2$\arcsec~(P. Garnavich, Private Communication).
There are H$_2$ images at 2.12\mum~of HH 7/11 obtained with higher angular resolution,
$\sim 0.\arcsec5-0.\arcsec 6$ (see Chrysostomou et al.~\citeyear{chrys00} for details)
which show an even more clear view of the outflow.
From Figure \ref{f1} we notice, (i) the different structure of HH 7,
the leading {\it Working Surface} (WS) of the jet, which at optical wavelengths shows two
well separated regions that have been identified with the bowshock and the Mach disk;
(ii) the absence of a compact counterpart in H$_2$ to the optical HH 11 knot; and (iii)
that despite these differences there is a `one to one' correlation between the main
``knots'' in [\ion{S}{2}] and H$_2$.

The high angular resolution NICMOS 2.12\mum~continuum 
subtracted image and a comparison of 
the H$_2$ emission with that of \Ha~and [\ion{S}{2}] from the WFPC2 camera,
are displayed in Figure \ref{f2ab}.
These images show that, in detail, the atomic and
NIR H$_2$ gases have more complex morphologies than those that
can be described in terms of simple ``bullets'' or ``knots''.
The gross properties of the emission from both gases have been
discussed in some detail by Hartigan, Curiel and Raymond (1989),
and so we will concentrate on two of the most interesting  
and relatively new features: the H$_2$ jet in SVS 13 and the HH 7 WS.

\subsection{H$_2$ Jet in SVS 13}

An enlargement of the NICMOS image around the SVS 13 source shows (Figure \ref{f3})
the presence of highly collimated and resolved H$_2$ emission, with a relatively high
flux density that we identified with a small jet.
The brightest component of the jet has 1.\arcsec24~in length 
at PA = 163\arcdeg~(although fainter emission extends up to 2.\arcsec4).
The NICMOS image therefore resolves and confirms that 
the 2.\arcsec6 asymmetric feature arising SE of SVS 13, detected
in recent H$_2$ Fabry-Perot (F-P) observations by Davis et al. (2002), 
corresponds indeed to a small-scale jet.
At a distance of 350pc, the jet has a length of 430 AU, making
it one of the smallest H$_2$ jets known.
The subtracted PSF SVS 13 image  has a small ''bump'' in the North aligned 
with the jet and could be tracing the counter-jet. 
The jet itself is narrower close to the source with an 
aspect ratio of $\sim 2.2$, and becomes wider at $\sim$ 0.\arcsec64 (222 AU)
from SVS 13. Considering that we are looking at this outflow in projection,
at approximately 30\arcdeg~with respect the plane of the sky
(Herbig \& Jones~\citeyear{her83}), the structure of this jet 
is remarkably similar to that of DG Tau,
the best example of an {\it optical} micro-jet 
(Solf \& B\"ohm~\citeyear{sb93}; Kepner et al.~\citeyear{kep93};
Lavalley et al.~\citeyear{lav97};
Lavalley-Fouquet, Cabrit \& Dougados~\citeyear{lav00};
Dougados et al.~\citeyear{cath00}).

The presence of an H$_2$ jet is consistent with 
other molecular tracers of high velocity gas. For example,
high-sensitivity interferometric CO J=2-1 observations show already
a ``bridge'' of gas between SVS 13 and the rest of the outflow, including
a jet-like feature in the `extremely-high-velocity' (EHV) gas near the source 
(Bachiller et al.~2000). The CO J=2-1 gas has a peak radial velocity of 
$\sim -172$ \kms, similar to that of $-175\pm 50$ \kms~observed in [S~II]
6717/31 in HH 11, the nearest knot to SVS 13 (Solf \& B\"ohm~\citeyear{sb87}). 
And finally, high spectroscopic resolution observations at 2.12\mum~by Davis
et al. (2001) clearly show two velocity components in the H$_2$ gas within 
1\arcsec~of the source, with a blueshifted high velocity component of 
$\sim$ 100 \kms.

One of the interesting things about this jet is that if one assumes
a flow velocity of 200 \kms~for HH 7/11, consistent with the velocity derived
from the optical proper motions (Noriega-Crespo \& Garnavich \citeyear{nori01}),
these parameters set a kinematical age of $\sim 10$ years. SVS 13 became an 
optically visible source during its last outburst in 1988-90 
(Eisl\"offel et al.~\citeyear{eis91}; Liseau, Lorenzetti \& Molinari~\citeyear{llm92}),
almost 10 years ago, and so is quite possible that the observed jet is the 
result of such energetic event.

We can compare the H$_2$ 2.12\mum~emission of the bowshock, Mach disk and 
jet, to have a idea of the {\it relative} physical conditions in these
regions. For example, the jet flux density $S_{j} = 4.1\pm 0.1\times 10^{-14}$ \fdens,
is three times higher than that of the  bowshock,
$S_{bs} = 1.4 \pm 0.1\times 10^{-14}$ \fdens, and ten times higher
than that of Mach disk, $S_{md} = 0.4\pm 0.1\times 10^{-14}$ \fdens. 
We can interpret these flux density ratios using the recent spectroscopic
results obtained in the far infrared (FIR) with the
Infrared Space Observatory (ISO) and molecular
shock models (Molinari et al. \citeyear{mol00}). 
HH 7 WS and SVS 13 have similar intensities in the 
(0-0) S(1)-S(7) H$_2$ pure rotational lines, as well as
in other shock indicators like [\ion{Ne}{2}] 12.8\mum~and [\ion{Si}{2}] 34.8\mum,
which suggests that a 20-30 \kms~molecular C-shock can explain 
the NIR and FIR H$_2$ emission in both objects (Molinari et al. \citeyear{mol00}).
On the other hand, molecular shock models at 20-30 \kms~for Hydrogen densities
in the range of $10^4 - 10^5$ \cc~predict that the line intensity of the
2.12\mum~line is $\sim 30$ times stronger in the high density models
(Kaufman \& Neufeld~\citeyear{kau96}).
So to satisfy both ISO observations and theoretical models: (i)
the observed flux density ratio ${S_{j}\over S_{bs}} = 3.1$ requires a higher 
density
for the jet, i.e. that the molecular jet is quite dense after leaving SVS 13, 
with a Hydrogen gas density of $few \times 10^4 - 10^5$ \cc~(Molinari et al.~
\citeyear{mol00}); and (ii) the bowshock is more dense ($\sim 10^4$ \cc~) 
than the Mach disk.

The presence of an H$_2$ jet ejected from this pre-main sequence 
source of such small angular scale, explicitly shows that 
contamination by an outflow in the search for
H$_2$ disk systems around young stars is possible, although perhaps not
as persistent as these systems become older. For example, the H$_2$ (0-0)
S(1) line flux around SVS 13 is $4.0\pm0.5 \times 10^{-13}$ \flux, i.e.
5-10 times higher than those found in the T Tauri and Debris-Disk samples of 
Thi et al. (2001), but certainly comparable to the values detected in
their Herbig Ae sample.

\subsection{HH 7 Working Surface}
\label{ws7}

HH 7 is the leading working surface of the HH 7/11 jet and
is  perhaps the most characteristic feature of the outflow. 
An enlargement of the groundbased observations
in Figure \ref{f4} shows that the H$_2$ emission spatially coexists with that 
from the atomic gas in both the arc-like bowshock and its Mach disk. 
Figure \ref{f4} also displays the NICMOS and WFPC2  observations of the same region, 
unfortunately the HST optical images are vignetted, one of the reason why they 
have not been published before, and so the emission corresponding to the 
bowshock is missing.
The bowshock and Mach disk are separated by $\sim 3.3$\arcsec~(1138.5 AU) 
with peak intensities in H$_2$ at $\alpha(2000) = 3^{h}29^{m}08.^{s}53$, 
$\delta(2000)$ = 31\arcdeg15\arcmin25.\arcsec5~
and $\alpha(2000) = 3^{h}29^{m}08^{s}.31$, 
$\delta(2000)$ = 31\arcdeg15\arcmin24.\arcsec2, respectively. 
The H$_2$ bowshock is large ($\sim$ 6\arcsec) and 
subtends a relatively smooth and complete arc of $\sim 140$\arcdeg. 
The fact that this shell is not fragmented suggests the presence 
of a sizeable magnetic field. Theoretical arguments and numerical 
simulations have shown that a magnetic field can modify the morphology of 
a jet (e.g. Frank et al. \citeyear{adam98}; Stone \& Hardee~\citeyear{stone00}),
and that a compressed magnetic field stabilizes a dense gas shell against
Rayleigh-Taylor instabilities if $B\sim 10^{-4}$ G (e.g. Blondin, Fryxell \& 
K\"onigl~\citeyear{blon90}). In our case, if the initial magnetic field is
given by $B_0 = {n_0}^{1/2} 10^{-6}$ G, where the preshock density,
is $n_0 = 10^4$ \cc, as deduced from the H$_2$ emission, then $B_0\sim 300$
$\mu$G. Therefore even a slow shock of $v_s$ = 20 \kms~(consistent with
the H$_2$ emission) can compress the magnetic field by a factor 15
($\sim 0.78~~v_s$[\kms],  McKee \& Hollenbach~\citeyear{mckee80}), 
i.~e. enough to meet the stability criteria.

In the NICMOS images the H$_2$ emission at the Mach disk is relatively faint
(see Figure \ref{f4}). The high-resolution H$_2$ groundbased observations 
by Chrysostomou et al. (2000) show a more clear picture of this feature.
The presence of a Mach disk in H$_2$ is perhaps more convincing if one uses,
for instance, the kinematical information available from F-P observations.
At optical and NIR wavelengths, is possible to distinguish
two distinct kinematical components corresponding to the bowshock and the
Mach disk separated by $\sim$ 3\arcsec~- 4\arcsec~
(Stapelfeldt~\citeyear{krs91}; Carr~\citeyear{carr93};
Davies et al.~\citeyear{chris01}, their Fig. 1).
The physical conditions at the Mach disk can be estimated in relative terms
by comparing its H$_2$ emission to that of the bowshock.
The flux density ratio between the bowshock and Mach disk is 3.5, and this 
could be due to a slight difference in density and/or shock velocity. 
Proper motion measurements set an upper limit of 30 \kms~for the velocity
of the Mach disk (Noriega-Crespo \& Garnavich~\citeyear{nori01}), 
which is very close to the predicted {\it shock velocity} of the entire
working surface. This means that even a complete thermalization
of the shock could not account for the observed ratio. If we
assume similar pre-shock conditions for the bowshock and the Mach disk,
and knowing that Mach disk velocity ($v_{md}$) depends on the jet ($v_j$), 
bowshock ($v_{bs}$) and pre-shock medium ($v_{med}$) velocities, i.~e.
$v_{md} = v_j - (v_{bs} + v_{med})$, then is likely that at the Mach disk 
the shock velocity itself is less than 20 \kms.

The question is whether or not the H$_2$ emission observed at
the Mach disk truly corresponds to that excited by a magnetic precursor 
(C-shock) or is the result of some other processes, 
like entrainment or a J-type shock.
The detailed answer to this question requires a MHD numerical simulation 
that is beyond the scope of this paper. The complex interaction
of the ion and electron fluids with a magnetic field, coupled with
the limited grid resolution of most numerical simulations, could explain
why an H$_2$ jet shock or Mach disk has not been seen in recent molecular 
jet models (e.~g. Raga et al.~\citeyear{raga95}; Suttner et al.~\citeyear{sutt97})).

One can ask, however, if the conditions at the working surface
are the necessary for the development of a magnetic precursor, 
capable to excite the H$_2$ in the Mach disk. Let's assume first 
a 200 \kms~jet flow that generates shock velocities $\sim 100$ \kms.
In the most simple scenario (without dust grains) a magnetic precursor 
needs molecular gas with a low ionization fraction and a moderate magnetic 
field (Draine \& McKee~\citeyear{dm93}). At the working surface,
the bowshock is the first shock that interacts with the surrounding gas, 
and although a magnetic precursor can excite the gas ahead of the shock, 
eventually (for shock velocities of $\sim 30-50$ \kms) the H$_2$ molecules 
will be dissociated (Smith~\citeyear{sm94}; Smith \& McLow~\citeyear{mmm97}),
except at the "wings" of the bowshock, where the shocks are oblique
and the molecular gas is only excited
(Smith~\citeyear{smd91}; Davis et al.~\citeyear{chris99}).
The emission of the atomic species, e.g. \Ha~and [\ion{S}{2}], 
in the HH7 WS indicates that the postshock gas is partially ionized 
and warm, and not ideal for a magnetic precursor, as the jet shock encounters 
this material. The exception could be again at the bowshock wings where 
the gas recombines more rapidly. This region is indeed where we detect 
the `strongest' H$_2$ emission at the position of the optical Mach disk.
Let's consider now a lower velocity flow such that the shock
velocity is $\le 50$ \kms. In this case we can refer to recent results 
from Lim, Rawling \& Williams (2001), where their adaptive grid 2D 
hydro-code incorporates the chemistry of 102 species, including those 
for molecules such as H$_2$O or H$_2$. In their three models (A-C), 
for different jet/environment density ratios, a H$_2$ jet shock is created.
For example in their model C, with $n_{jet} = 50$ \cc~and $n_{env}=5$ \cc, 
the highest H$_2$ density occurs at the Mach disk
(Lim, Rawling \& Williams~\citeyear{lim01}; their figure 6).

The above scenarios for the H$_2$ emission at the Mach disk suggest 
either a fine tuning of the physical conditions of the postshock 
gas at the working surface to drive a magnetic precursor, or the need 
to incorporate in the hydrodynamics the entire chemistry network.
Perhaps a more straightforward explanation is that a fraction of 
the H$_2$ gas has been entrained by the jet in its interaction 
with the molecular ambient medium. This scenario has been recently explored 
by Raga et al. (2002), where an atomic jet collides sideways with
a molecular H$_2$ cloud (see their Fig 1), and as the jet ``bounces''
from the cloud entrains some of the molecular gas. In practice,
there is not need of such strong interaction and is enough for the jet
to strike some dense molecular gas along its path, as has been
suggested previously for this system (see e.~g. Knee \& Sandell~
\citeyear{ks00}; Sandell \& Knee~\citeyear{sk01}).

\subsection{HH 8-11}

We have described in the previous two sections what we 
consider the most outstanding features observed with the high angular 
resolution NICMOS observations. In this section, we describe briefly the 
main characteristics in of the other knots HH 8-11 
in the H$_2$ 2.12\mum~and [\ion{S}{2}] 6717/31 lines (Figure \ref{f5}).

HH~8. In the WFPC2 images this knot has a filamentary structure
that spans $\sim$ 7\arcsec, with a bright small core ($\sim$ 2\arcsec)
where both the H$_2$ and atomic emission coincide,
although with distinct morphologies (Figure \ref{f5}).
Kinematically, the knot follows the overall flow pattern
defined by HH 7 and HH 11 (Herbig \& Jones~\citeyear{her83}; 
Noriega-Crespo \& Garnavich~\citeyear{nori01}), however,
the H$_2$ F-P observations show the molecular gas with a radial
velocity of $\sim 40$ \kms~slower than the [S~II] gas and with a 
component at zero and positive velocities (Carr~\citeyear{carr93}). 
This suggests that the H$_2$ emission could be either arising from behind 
the optical object, and is observed in projection (as expected from
a bowshock, with a  weak shock component at its tail); 
or could be the result of entrainment at the edge of the cavity created 
by the jet, but seen also in projection. In both situations,
the H$_2$ gas would have a lower velocity than the atomic/ionic gas.

HH 9. This knot is not visible in the NICMOS image, which is not
a surprise given the short exposure and considering that even in
groundbased observations is hard to detect (e.~g. Figure \ref{f1}).
At optical wavelengths, the WFPC2 images show a `fuzzy'
knot of $\sim$ 6\arcsec~in size; the knot
has a [\ion{S}{2}]/\Ha~ratio of $\sim 1.1$
indicating its low excitation (Table 2).
HH 9 has a very low radial velocity and essentially
zero proper motion (Solf \& B\"ohm~\citeyear{sb87}; 
Noriega-Crespo \& Garnavich~\citeyear{nori01}),
so could be either part of the backflow of
the jet envelope or an ambient medium condensation
that has been excited by the outflow.

HH 10. Already from the groundbased optical images was possible
to infer a complex morphology for this ``knot'' (Figure \ref{f1}),
but certainly not with the detail of the WFPC2 images.
The morphology of the [\ion{S}{2}] and \Ha~gas emission are
quite similar in shape and span some $\sim$ 14\arcsec~in the N-S direction
while the H$_2$  emission, as in the case of HH 8, is not as extended and
well defined (Figure \ref{f5}).

HH 11. This object is one of the few condensations in the HH 7/11 
chain, that at optical wavelengths, looks like a ``bullet''. 
HH 11 is the fastest moving knot of the outflow, with a space 
velocity in the atomic/ionic gas of 190 \kms~
(Noriega-Crespo \& Garnavich~\citeyear{nori01}), and
has the highest excitation in the flow (with a [\ion{S}{2}]/\Ha~
ratio of $\sim 1.8$), indicating shocks as strong as $\sim$ 60 \kms.
Situated at $\sim 10$\arcsec~from SVS 13, this condensation
was probably ejected $\sim 87$ yr ago. These properties reinforce 
the idea that we are witnessing a jet interacting
with a moving medium set in motion by a previous ejection 
event (Raga et al.~\citeyear{raga90}).
The observed H$_2$ emission is located further back from 
the main object trailing behind the atomic/ionic gas
(defined by [\ion{S}{2}] and \Ha~emission),
as one expects from the most simple bowshock models
which include H$_2$ emission (Smith~\citeyear{smd91}).
Deep groundbased NIR images at 2.12\mum~show a bit more extended
emission in HH 11, but still predominantly behind the optical 
counterpart (e.~g. Chrysostomou et al. \citeyear{chrys00}).
Therefore, HH 11 is most likely to be one of those rare cases 
where the H$_2$ molecules are dissociated near the stagnation 
region of the bowshock, but they survive and are excited
in those regions where the shocks become weaker (oblique), 
i.e. at the bowshock wings (Smith~\citeyear{smd91}).

SVS13 NE. WFPC2 images in [\ion{S}{2}] and \Ha~show
a small condensation at approximately 5\arcsec~NE of SVS 13,
about $\sim 3\arcsec$ in size (Figure \ref{f5}), 
that seems to trace the starting point of an arc-shaped structure 
north and parallel to the HH 7/11 outflow (see Figure \ref{f1}).
Presently, we don't have any kinematical information on this
knot. But the overall structure reminds us of another well known
outflow in L1551, where it has been suggested that the
opening angle has changed as a function of time, going
from a broad to a narrow configuration (Davis et al.~\citeyear{chris95}
their figure 9), creating in the process a fan shaped cavity.

\section{Conclusions}
\label{concl}

We have resolved two remarkable features in the 
molecular Hydrogen emission of the HH 7/11 outflow
thanks to HST NICMOS images at 2.12\mum:
a jet with a length of 430 AU arising from the 
SVS 13 driving source, and the Mach disk in HH 7, leading 
working surface. These observations strongly support the 
presence of small-scale H$_2$ jets arising from Class I/O 
sources (Davis et al.~\citeyear{chris02}), and open
the possibility that a jet shock can be detected in H$_2$.

Using previously published infrared spectroscopic observations,
coupled with molecular shock models, we have determined that:
(i) the jet can have a density as high as $10^5$ \cc, for
shock velocities of 20-30 \kms, (ii) the magnetic field plays
a major role in stabilizing the HH 7 bowshock, (iii)
the Mach disk H$_2$ emission is probably produced by shocks 
of less than 20 \kms; and (iv)
the complex distribution of the atomic and molecular gases
in HH 7/11, depicted by the NICMOS and WFPC2 images,
coupled with its kinematics, suggests a strong 
interaction of this outflow with its circumstellar medium.

\begin{center} {\it Acknowledgments} \end{center}
A.N-C. research is partially supported by NASA through a contract with
Jet Propulsion Laboratory, California Institute of Technology,
and ADP Grant NRA0001-ADP-096. We thank A. Moro-Mart{\'{\i}}n
for critical and useful comments.

\clearpage

%\makeatletter
%\def\jnl@aj{AJ}
%\ifx\revtex@jnl\jnl@aj\let\tablebreak=\nl\fi
%\makeatother
%\ptlandscape  

\begin{deluxetable}{llcll}
\tablecolumns{10}
\tablewidth{400pt}
\tablecaption{HH 7/11 Data} 
\tablehead{
\colhead{Instrument} & \colhead{Camera} &
\colhead{Date} & \colhead{Filter} & \colhead{Exposure(sec)}}
\startdata
NICMOS &  NIC2   & 09 Jan 98 & F187N, F204, F212N & 80, 80, 160 \\
WFPC2  & \nodata & 18 Oct 98 & F656N, F673N & 3600, 3600 \\
\enddata
\end{deluxetable}

\clearpage

\begin{deluxetable}{l|ccc|ccc|ccc}
\tablecolumns{10}
\tablewidth{530pt}
\tablecaption{Fluxes by Region} 
\tablehead{
\colhead{}&\multicolumn{3}{c}{H$_2$ (F212N)}&\multicolumn{3}{c}{H$\alpha$ 
(F656N)}&\multicolumn{3}{c}{[(\ion{S}{2})]} \\
\colhead{}
&\colhead{Flux\tablenotemark{a}}
&\colhead{Area\tablenotemark{b}}
&\colhead{Flux/Area\tablenotemark{c}}
&\colhead{Flux\tablenotemark{a}}
&\colhead{Area\tablenotemark{b}}
&\colhead{Flux/Area\tablenotemark{c}}
&\colhead{Flux\tablenotemark{a}}
&\colhead{Area\tablenotemark{b}}
&\colhead{Flux/Area\tablenotemark{c}} \\
\colhead{Region}
&\colhead{($\times$10$^{-14}$)}&\colhead{}&\colhead{($\times$10$^{-15}$)}
&\colhead{($\times$10$^{-18}$)}&\colhead{}&\colhead{($\times$10$^{-19}$)}
&\colhead{($\times$10$^{-18}$)}&\colhead{}&\colhead{($\times$10$^{-19}$)}
}
\startdata
HH 7 &36.3$\pm$0.2 &50.1 &7.3 &4.78$\pm$0.01 &80.6 &0.59 &13.74$\pm$0.02 &80.6 &1.7 \\
HH 7\tablenotemark{d}&$\cdots$&$\cdots$&$\cdots$&0.74$\pm$0.01&4.2&1.76&2.85$\pm$0.01&4.2
&6.8\\
HH 8 &6.2$\pm$0.2  &30.0 &2.1 &3.12$\pm$0.01 &26.5 &1.18 &2.81$\pm$0.02  &26.5 &1.1 \\
HH 9 &$\cdots$& $\cdots$ & $\cdots$ & 0.84$\pm$0.01 & 19.6 & 0.43& 3.53$\pm$0.01 & 19.6 & 1.8 \\
HH 10&6.4$\pm$0.2  &36.8 &1.8 &3.11$\pm$0.01 &40.9 &0.76 &8.24$\pm$0.02  &40.9 &2.0 \\
HH 11&3.0$\pm$0.2  &22.7 &1.3 &2.91$\pm$0.01 &12.2 &2.39 &5.27$\pm$0.01  &12.2 &4.3 \\ 
\tableline
\enddata
\tablenotetext{a}{erg s$^{-1}$ cm$^{-2}$}
\tablenotetext{b}{arcsec$^{-2}$; errors for the derived area are $\pm$0.02}
\tablenotetext{c}{erg s$^{-1}$ cm$^{-2}$ arcsec$^{-2}$}
\tablenotetext{d}{Data for the F673N and F656N emission associated with a small 
area encompassing just the Mach disk}

\end{deluxetable}

\clearpage
%figure captions

\figcaption[f1.eps]{\label{f1}
Groundbased images of HH 7/11 outflow (at the same scale) in 
[\ion{S}{2}] 6717/31 \AA~(top)
and H$_2$ 2.12\mum~(bottom). There is a one-to-one correspondence for each
of the main objects, even for HH 9 which is quite faint in H$_2$.}

\figcaption[f2a.eps]{\label{f2ab}
(top) NICMOS continuum subtracted grayscale image at H$_2$ 2.121\mum~(F212N) of 
the HH 7/11 outflow. (bottom) Combined false three color image of HH 7/11.
Red is F212N (H$_2$), green is F673N ([\ion{S}{2}] 6717/31 \AA),
and blue is F656N (H$\alpha$).}

\figcaption[f3.eps]{\label{f3}
Continuum and PSF subtracted grayscale F212N
image of SVS 13 and its jet arising at the South.
The overlayed F212N contours correspond to 1.2,1.7,2.2,2.8,3.4,3.9,9.5,
and 15.1$\times10^{-19}$ W m$^{-2}$}

\figcaption[f4.eps]{\label{f4} (left) A comparison of the
groundbased observations of HH 7 working surface in
H$_2$ at 2.12\mum~(grayscale) and [\ion{S}{2}] 6717/31 \AA~
(white contours), which shows the spatial coexistence of
the atomic and molecular emission. (right) 
A contour map of NICMOS and WFPC2 observations
of HH 7, with  H$_2$ at 2.12\mum~(black) and 
[S~II] 6717/31 \AA~(red). The sharp edge of the [S~II] 
emission is due to vignetting of the WFPC2 image (see Fig. 1). 
The location bowshock and Mach disk
are marked. H$_2$ contour levels are 0.3, 0.5, 0.7, 0.8,
1.1, 1.3, 1.5, 1.7, 1.9, 2.1, 2.2 and 2.5 $\times10^{-19}$ W m$^{-2}$.}

\figcaption[f5.eps]{\label{f5}In all 4 figures 
black contours are H$_2$ (F212N), red are [\ion{S}{2}] (F673N),
and the data have been boxcar smoothed with a 3$\times$3 pixel 
kernel size to improve their quality. (From the top left)
{\bf HH 8 map}.
Contour levels for H$_2$ are 1.5, 2.2, 2.9, 3.6,
4.4, 5.1, 6.5, 7.8, 9.2, 10.5, 11.9, 13.3, 14.6 and 16.0 $\times10^{-20}$ W m$^{-2}$.
Contour levels for [\ion{S}{2}] are 0.67, 0.84, 1.0, 1.2, 1.35, 1.5, 1.7, 1.85, 
2.3, 2.0, 7.0, 3.1, 3.5, 3.95, 4.4 and 4.8 $\times10^{-26}$ W m$^{-2}$.
{\bf HH 10 map}. Contour levels for H$_2$ are 1.5, 1.9, 2.3, 2.7, 3.0,
3.4, 3.8, 4.2, 4.6, 4.9 and 5.3 $\times10^{-20}$ W m$^{-2}$. Contour levels for
[(\ion{S}{2})] are 0.88, 1.2, 1.6, 1.9, 2.3, 2.6, 3.0, 3.3, 3.7, 4.0, 4.4, 4.7 
and 5.1 $\times10^{-26}$ W m$^{-2}$.  
{\bf HH 11 map}. Contour levels for H$_2$ are 0.6, 1.0, 1.5, 1.9, 2.3, 3.0,
3.7, 4.4, 5.1 and 5.8 $\times10^{-20}$ W m$^{-2}$. Contour levels for
[\ion{S}{2}] are 0.4, 0.8, 1.1, 1.5, 1.8, 2.2, 2.5, 2.8, 3.2, 3.5,3 .9, 4.2, 4.6 
and 4.9 $\times10^{-26}$ W m$^{-2}$.  
{\bf NE of SVS13 map}. Contour levels range 
from 0.3 to 44$\times10^{-26}$ W m$^{-2}$
for \Ha~and 0.1 to 26$\times10^{-26}$ W m$^{-2}$  for [\ion{S}{2}], 
where the largest values pertain to the brightest contours of SVS 13.}

\end{document}